\documentstyle[11pt,paspconf,epsf]{article}

\begin{document}

\title{Metallicity calibration of theoretical stellar SEDs using
$UBVRIJHKL$ photometry of globular clusters}
\author{P. Westera, Th. Lejeune and R. Buser}
\affil{Astronomisches Institut der Universit\"at Basel, Switzerland}

\begin{abstract}
Although stellar SED models are constantly improving and seem to show similar
differential tendencies as measured spectra, their absolute properties
(eg. absolute flux in different wavelength regions) can still differ
significantly from the measured ones. Therefore, it is necessary to
``color-calibrate'' the theoretical spectra to obtain realistic model-based
stellar spectra, which can then be used for many purposes, such as population
synthesis models. \\
In order to calibrate a hybrid library (consisting of the Kurucz -, the Allard
and Hauschildt -, and the Scholz-models) of theoretical SEDs, $UBVRIJHKLM$
photometric data of solar neighborhood stars were collected from the literature,
and the solar metallicity models were calibrated using an algorithm developed by
Cuisinier et al. The corrections were then propagated through the whole grid of
models, resulting in the ``semi-empirical'' models (available on ftp from the
university of Basel). \\
In order to extend the calibration to non-solar abundances, we have collected
$UBVRIJHKL$ photometric data from well-known Galactic globular clusters, covering
a wide range of metallicities. \\
We shall present first results, using the metallicity- calibrated library for
calculating isochrones of metal-poor clusters. In particular, the observed slope
and curvature of the RGB appears to be reproducible in a systematically correct
manner for the first time.
\end{abstract}

\keywords{UBVRIJHKL photometry,globular clusters,SED models,calibration,metallicity}

\section{Introduction}

As present grids of theoretical stellar spectra suffer from intrinsic inhomogeneities and incompleteness and show large systematic discrepancies with empirical calibrations due to unavailable molecular opacity, we have undertaken the construction of a comprehensive combined library of {\em realistic} stellar flux distributions. Empirical ${T_{\rm eff}}$-color relations in $UBVRIJHKL$ photometry are used to calibrate the spectra using an algorithm developed by Cuisinier et al. (1995). \\
In a first step, the calibration was carried out for solar metallicity, and the differential properties with metallicity were preserved (Lejeune et al. 1997). As expected, some discrepancies at low ${\rm [Fe/H]}$ remained, so the next step was to collect photometry of Galactic globular clusters in order to do the calibration in a metallicity-dependent way. \\
Although much work still has to be done, first results are presented here.

\section{The semi-empirical (BaSeL 3) SED library}

In order to cover the necessary range of parameters, we combined the Kurucz (1995) models ($3500 K \leq {T_{\rm eff}} \leq 50000 K, -5.0 \leq {\rm [Fe/H]} \leq +1.0, 0.0 \leq \log g \leq 5.0$) with the Scholz (1997) models ($2500 K \leq {T_{\rm eff}} \leq 3500 K, -2.0 \leq {\rm [Fe/H]} \leq +0.5, -1.02 \leq \log g \leq 2.0$) and the Allard \& Hauschildt (1995) models ($2000 K \leq {T_{\rm eff}} \leq 3500 K, -4.5 \leq {\rm [Fe/H]} \leq +0.5, 3.5 \leq \log g \leq 5.5$). \\
This grid of models was then calibrated using $UBVRIJHKLM$ photometric data of solar neighbourhood stars (dwarfs and giants) and an algorithm developed by Cuisinier et al. (1995, see Lejeune et al. 1998), which modifies the model spectra to match the calibration colors for solar abundances and then preserves the differential properties with metallicity and surface gravity of the original models. \\
As expected, this so-called semi-empirical library still has its limitations at low metallicities. Although the tendencies with decreasing ${\rm [Fe/H]}$ seem to be well reproduced for visible to near infrared colors ($B$-$V$, $V$-$I$, $V$-$K$), the effect of metallicity in the far-infrared ($J$-$H$, $H$-$K$, $J$-$K$, $K$-$L$) and the ultraviolet ($U$-$B$) are less well reproduced. This was expected from the fact that some important molecular opacity data are still missing. \\
The semi-empirical (BaSeL 3) SED library is currently available by ftp from the Astronomical Institute of the University of Basel, Switzerland.

\section{The calibration data}

In order to overcome these shortcomings, we performed a metallicity-dependent calibration of the grid, using empirical photometry from the literature (see the reference list) of the best-studied Galactic globular clusters with metallicities down to -2.16 (${\rm [Fe/H]}$-values from Carretta \& Gratton 1997), as globulars are abundant in stars of known metallicity and all evolutionary stages. \\
Combined multi-color ($UBVRIJHKL$) - ${\rm [Fe/H]}$ - ${T_{\rm eff}}$ - $\log g$ relations were synthesised using $E_{B-V}$ and $(m$-$M)_{V}$ values from the Harris online catalog of globular cluster parameters, the ${T_{\rm eff}}$ -  $V$-$K$ relation from Ridgway et al. (1980) and  empirical ${T_{\rm eff}}$ - $\log g$ relations for red giants from Frogel, Persson and Cohen (1978, 1981 and 1983, ${\rm [Fe/H]}$-dependent) and dwarfs from Angelov (1996, ${\rm [Fe/H]}$-independent), where the large gaps were filled using the differential properties (in ${T_{\rm eff}}$) of the semi-empirical grid.

\section{First results}

At present, there has only been time to produce temperature-color relations for ``RGB''s (the models evaluated for the parameter values of empirical RGBs). These first results are presented in the following figure, wherein the new (metallicity-calibrated) models (dashed) are compared with the BaSeL 3 (semi-empirical) models (dotted) and the empirical relations from the clusters that were used for this calibration and solar metallicity (solid).
As expected, the changes in most colors are relatively minor, but the new grid appears to show smoother relations, and improvements can be seen for $U$-$B$, $J$-$H$, $H$-$K$ and $J$-$K$, especially at low temperatures.
\begin{figure}[!ht]
\plotfiddle{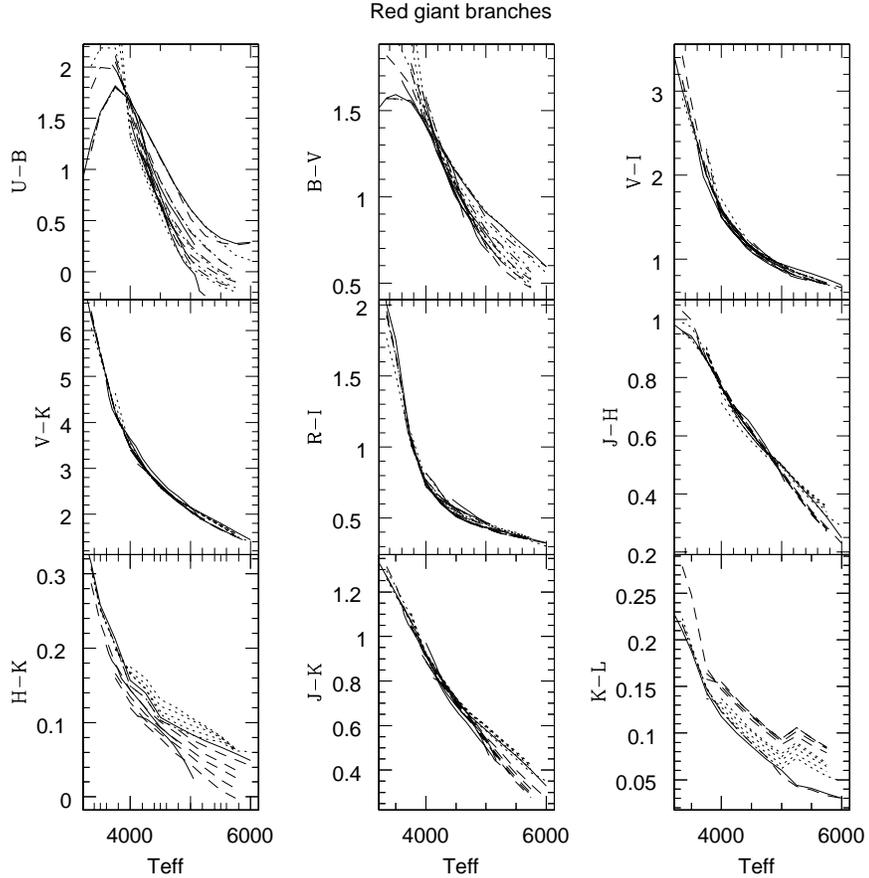}{11.0cm}{0}{60}{60}{-190}{-105}
\caption{Temperature-color-relations for the metallicity-calibrated models (dashed), the BaSeL 3 models (dotted), and the clusters 47 Tuc, M 5, M 3, NGC 6397 and M 92 and solar metallicity (solid).}
\end{figure}

\section{Conclusions}

Although more relevant tests still have to be performed, some small improvements seem to have been achieved with this first metallicity-calibration of the Basel library of stellar SEDs. The real breakthrough however, can only be achieved after the huge gaps in the empirical data are filled. In particular, $U(BV)R(I)JHKL$ photometry needs to be performed for globular clusters of low ${\rm [Fe/H]}$ down to the main sequence. \\
We expect to obtain new observations by using the ESO NTT / 3.6 m telescopes soon. In a first run, we plan to collect $UBVRI$ photometry data for Galactic globular cluster stars covering luminosities from the tip of the RGB down to the main sequence at $M_{V} \sim 8^{m}$, and metallicities in the range $-2.1 < {\rm [Fe/H]} < -1.2$.

\acknowledgments

I'd like to thank Gisela Branco for her graphical support to the poster.


\begin{references}
\reference Alcaino, G., \& Liller, W. 1986, \aj, 91, 87
\reference Alcaino, G., Liller, W., Alvarado, F., Kravtsov, V., Ipatov, A., Samus, N., \& Smirnov, O. 1997, \aj, 114, 1067
\reference Allard, F., \& Hauschildt, P. H. 1995, \apj, 445, 433
\reference Angelov, T. 1996, BOBeo, 153, 19
\reference Arribas, S., \& Martinez Roger, C. 1987, \aap, 178, 106
\reference Bessell, M. S., Castelli, F., \& Plez, B. 1998, \aap, 333, 231 
\reference Bica, E., Ortolani, S., \& Barbuy, B. 1994, \aap, 106, 161
\reference von Braun, K., Chiboucas, K., Minske, J. C., Salgado, J. F., \& Worthey, G. 1998, \pasp, 110, 810V
\reference Carretta, E., \& Gratton, R. G. 1997, \aaps, 121, 95
\reference Cohen, J. G., Frogel, J. A., \& Persson, S. E. 1978, \apj, 222, 165
\reference Cuisinier, F., Lejeune, T., \& Buser, R. 1995, Proc. of the IAU Symp., 171, 355
\reference daCosta, G. S., \& Armandroff, T. E. 1990, \aj, 100, 162
\reference Drissen, L., \& Shara, M. M. 1998, \aj, 115, 725
\reference Ferraro, F. R., Carretta, E., Corsi, C. E., Fusi Pecci, F., Cacciari, C., Buonanno, R., Paltrinieri, B., \& Hamilton, D. 1997, \aap, 320, 757
\reference Frogel, J. A., Persson, S. E., \& Cohen, J. G. 1981, \apj, 246, 842
\reference Frogel, J. A. , Persson, S. E., \& Cohen, J. G. 1983, \apj, 53, 713
\reference Grebel, E. K., \& Roberts, W. J. 1995, \aaps, 109, 293
\reference Harris, W. E. 1997, Catalog of parameters for Milky Way globular clusters, http://physun.physics.mcmaster.ca/GC/
\reference Hesser, J. E., Harris, W. E., VandenBerg, D. A., Allwright, J. W. B., Shott, P., \& Stetson, P. B. 1987, \pasp, 99, 739
\reference Kaluzny, J. 1997, \aaps, 122, 1
\reference Kaluzny, J., Wysocka, A., Stanek, K. Z., \& Krxeminsky, W. 1998, AcA, 48, 439
\reference King, I. R., Anderson, J., Cool, A. M., \& Piotto, G. 1998, \apj, 492L, 37
\reference Kurucz, R. L. 1995 (private communication)
\reference Lee, S-G., Lee, M. G., \& Kim, E. 1996, JKAS, 29, 171
\reference Lejeune, T.,  Cuisinier, F., \& Buser, R. 1997, \aaps, 125, 229
\reference Lejeune, T.,  Cuisinier, F., \& Buser, R. 1998, \aaps, 130, 65
\reference Montegriffo P., Ferraro F. R., Fusi Pecci F., Origilia L. 1995, \mnras, 276, 739
\reference Piotto, G., Cool, A. M., \& King, I. R. 1997, \aj, 113, 1345
\reference Ridgway, S. T., Joyce, R. R., White, N. M., \& Wing, R. F. 1980, \apj, 235, 126
\reference Rieke, G. H., \& Lebofsky, M. J. 1985, \apj, 288, 618
\reference Salaris, M., Chieffi, A., \& Straniero O. 1993, \apj, 414, 580
\reference Sandage, A. 1970 , \apj, 162, 841
\reference Sandquist, E. L., Bolte, M., Stetson, P. B., \& Hesser, J. E. 1996, \apj, 470, 910
\reference Scholz, M. 1997 (private communication)
\reference Stetson, P. B., \& Harris, W. E. 1988, \aj, 96, 909
\end{references}
\end{document}